# A Competition-based Pricing Strategy in Cloud Markets using Regret Minimization Techniques

S.Ghasemi[*], M.R.Meybodi[**], M.Dehghan[**], A.M.Rahmani[*]

[*]Department of Computer Engineering, Science and Research Branch, Islamic Azad University, Tehran, Iran
[**]Computer Engineering and Information Technology, Amirkabir University of Technology, Tehran, Iran.

*Abstract*— Cloud computing as a fairly new commercial paradigm, widely investigated by different researchers, already has a great range of challenges. Pricing is a major problem in Cloud computing marketplace; as providers are competing to attract more customers without knowing the pricing policies of each other. To overcome this lack of knowledge, we model their competition by an incomplete-information game. Considering the issue, this work proposes a pricing policy related to the regret minimization algorithm and applies it to the considered incomplete-information game. Based on the competition based marketplace of the Cloud, providers update the distribution of their strategies using the experienced regret. The idea of iteratively applying the algorithm for updating probabilities of strategies causes the regret get minimized faster. The experimental results show much more increase in profits of the providers in comparison with other pricing policies. Besides, the efficiency of a variety of regret minimization techniques in a simulated marketplace of Cloud are discussed which have not been observed in the studied literature. Moreover, return on investment of providers in considered organizations is studied and promising results appeared.

1. INTRODUCTION

Cloud Computing, introduced in the early 21$^{st}$ century, has emerged as one of the most attractive technologies in IT industry [1, 2, 3] since 2007. One of the main goals of this new computing paradigm is delivery of requested applications through provisioned virtualized resources [1, 4]. Recently, a great number of users has taken the advantages of services offered by Cloud computing providers. These providers are generally categorized into three groups: Infrastructure as a Service (IaaS), Platform as a Service (PaaS), and Software as a Service (SaaS) [1]. IaaS providers prepare computing and storage resources [5, 2] in form of virtual machines (VM); in addition to other providers, industrial and scientific organizations may utilize these services to run their applications [2]. SaaS providers offer applications to users without needs of infrastructures maintenance or development [6]. Existence of different providers offering their own services to potential users, makes the Cloud computing environment a competitive marketplace [7]. There are many studies that consider competition-based interactions between IaaS providers [3, 6, 8-16], between SaaS providers and IaaS providers or users and IaaS/SaaS providers [2, 4, 5, 9, 17-24], among SaaS providers [7, 14, 17, 25, 26]. In such investigations, Cloud is considered as a marketplace, including multi-tier economic interactions of SaaS, PaaS and IaaS providers and users which may strategically compete with each other within some tiers of this marketplace.

This study focuses on interactions of SaaS providers and users in form of a game, like [7, 14, 26]. Proposed method of [14] formulated a non-cooperative stochastic game as a Markov Decision Process (MDP), with a Markov Perfect Equilibrium solution. MDPs operate on completely visible environments while our considered environment is not completely visible. SaaS providers should attract users to increase their profit. Pricing strategies of provided applications have a major effect on their profit; initial cost of applications, resource appropriation costs, user willingness to pay, multi-tenancy of application and so forth are some of the most prominent factors of pricing strategies, studied in form of a complete information game in [7]. SaaS providers consider services based on their changing utilities, user's service choice behavior, and the like. [26] uses a non-cooperative game to study such service provisioning problem in multi-Cloud competitions in a way to obtain the largest market share. Besides, there are some other studies such as [27, 28] which examine the pricing challenges of applications; application pricing in a duopoly is investigated under some constraints in [29], using an incomplete information game, where players are SaaS providers whose strategies are pricing policies, regarding the main features of application prices. Users usually tend to use an application which has satisfying performance with the least price. Unlike [7, 14], the lack of information in some environments [17, 29-31], reaching an equilibrium seems to be out-of-the-way, and applying learning techniques facilitate the issue. The behavior of SaaS providers is studied [19, 23] to optimize costs of running tasks on virtual resources while they are competing with each other to use the best resources stored in IaaS providers, regardless of pricing policies.

However, most existing competition based mechanisms are inefficient and do not consider the application pricing strategies, regarding the competition between SaaS providers. In the present research, a competition based pricing mechanism is introduced which investigates various factors impacting the price of an application including development and deployment requirements. An efficient solution of an incomplete information game is studied based on regret minimization technique, using the updating equations to minimize the regret. Applying regret minimization techniques, instead of profit optimization ones, facilitates the solution as regrets may model the behaviors of players while their opponents perform more efficiently. Our proposed approach is applied to a pricing optimization game in which players are not aware of strategies of others.

The main features that differentiate our work from the studied works can be outlined as follows. First, the proposed game is between SaaS providers offering their provided applications to users in an incomplete information environment. Quantitative modeling of SaaS providers and internal and external validation of the model are some positive points of our study. Second, our paper covers some considerations in dynamic pricing of applications, whose prices have significant economical influences on decisions of SaaS providers. Third, as our considered game is an incomplete information one, a new regret minimization technique is applied to achieve the solution concept known as correlated equilibrium. Finally, we evaluate the performance of different regret minimization techniques, including external, internal and swap regret minimization techniques applied to our proposed approach. Moreover, the return of investment is studied for providers using the proposed pricing policies.

The rest of this paper is organized as follows. The review of the related works is presented in Section2. The system model, the problem statement and the related concepts of market state can be found in Section3. Section4 studies the formulated problem of the providers' pricing policy optimization, the game theory based model of Cloud application pricing, and the solution concepts of the proposed game. Section5 discusses the introduced experimental results, and demonstrates the efficiency of our proposed solution. The paper is ended with some concluding remarks in Section6.

## 2. RELATED WORKS

This section provides a comprehensive review of literature on economical profits of Cloud computing paradigm. There are various sorts of resources such as processors, a cluster of computing devices, storage space, online tools, data, information, and applications which make the provisioning process of Cloud requests much complex.

An auction is a primary market mechanism of selling commodities. In an auction based resource allocation, all parties do their best to improve their profits in the competition based marketplace. In [4], a taxonomy of the auction based resource allocation mechanisms in Cloud is provided; four major mechanisms are named 1) One-sided auctions consisting of three types as Reverse auctions, First-price sealed-bid auctions and Second-price sealed-bid auctions, 2) Double-sided auctions which has Continuous-time auctions, 3) Combinatorial auctions and 4) Other types of auctions. Auctions are mostly formulated as games. usage of game theory in auction of resource provisioning and pricing of Cloud are widely studied [3, 5, 6, 12, 13, 15, 20, 21, 23, 32, 10]. To maximize the profit of the providers, in [32], a Stackelberg game is assumed between operator of Cloud based integrated terrestrial satellite networks (connection of both satellite and RRHs to the Cloud for centralized processing) and users. The evolutionary game of users is solved with replicator dynamics to minimize energy consumption. Since there exist two types of users, when an evolutionary stable strategy of users is achieved the selections of both types of users are obtained for any pricing strategy. An optimal pricing strategy, that maximizes the profit of the operator and the Stackelberg equilibrium point of the system, is proposed by analyzing the evolution relationship between the profit and the pricing strategy. Thereafter, based on the optimal pricing strategy an optimal resource allocation scheme is obtained. [10] is another game theoretic approach to deal with pricing in resource allocation of Cloud providers. By using auction based allocation and pricing mechanisms, the satisfaction of both parties grows. The highest bids influence Cloud resources and services in a way that winners of the auction achieve the demanded quality. The major aim of the proposed winner-bid auction game is to increase the profits of the provider and user from different criteria. The game seals users' bids by a multi-criteria valuation-based bid function and sends them to the auctioneer. The determined Nash equilibrium specifies the prices to be paid by users in various cases and proves the truthfulness of the method. To analyze the reasonable allocation of resources, a Bayesian Nash equilibrium allocation algorithm is proposed in [3] which considers the heterogeneous distribution of resources, rational exchange behaviors of Cloud users, incomplete information and dynamic successive allocation. Cloud users reach Nash equilibrium allocation solutions by gambling stage by stage. Valerio et al. [5] solve the allocation problem of Cloud by proposing a two stage provisioning scheme; the first stage determines the number of required flat and on demand instances using standard optimization techniques, then in second stage, the competition of providers commences. The competition is among both IaaS providers and SaaS providers. The competition of providers in second stage is modeled by a Stackelberg game and the solution is achieved by solving a mathematical program with equilibrium constraints problem.

There are a variety of research that focused on pricing and resource management in Cloud computing [8, 12, 13, 15, 16, 19, 20, 21, 23, 26]. Effective pricing policies for a Cloud provider are discussed in [8]. Then, the offered prices are sent to users in form of 'take it or leave it' to decide whether to rent the Cloud resources or not. The goal of the research is to maximize the social welfare (the aggregate gain of the Cloud provider and Cloud users) that might not bring about maximization of an individual provider revenue. For both private Cloud and public Cloud, maximizing social welfare is more relevant than maximizing a provider revenue. It is to be noted that resource allocation mechanisms may be related to users' urgency and priorities. Zhang et al. propose pricing functions based on dual price design in competitive online algorithms related to the classic primal-dual framework [8]; their pricing function, bounded to total demand, is related to the online knapsack problem. In [7], some efficient pricing strategies are introduced for providers which bring about gaining leverage to retain user by making its switching costs too high, and to achieve high profitability by implementing the appropriate strategies.

Learning and heuristic algorithms are widely applied in the discussed problem of service provisioning in Cloud computing. [33] introduces a learning automata based resource provisioning approach that achieves cost-effective guarantees of demands which performs based on services of comprised applications and considers both cost and service requirements together for deploying applications. Another considerable learning based approach is introduced in [18], where the application of deep reinforcement learning is studied to obtain an optimal allocation/pricing policy that maximizes profit of the Cloud provider through trial and error. Du et al. [18] combine long short-term memory units with fully-connected neural networks to tackle online user arrivals, and adjust the output and update the methods to address both resource allocation and pricing. Moreover, Alsarhan et al. [9] use a learning based price control parameter in the proposed Cloud computing SLA framework to meet quality of service (QoS) demands in the market. To find an adaptable VM hiring policy to system changes including service cost, system capacity, and demand for service, a reinforcement learning approach is applied in the framework. The approach integrates computing resources adaptation with service admission control based on the reinforcement learning model. Besides, researchers in [2] analyses the efficiency of pricing process in a competing Cloud environment. The problem is modeled as a Markov game that Minimax-Q and Q-learning algorithms are applied to pricing policies of Cloud providers. In comparison of these two algorithms they find minimax-Q more suitable for analyzing the competition and adhering Cloud users while Q-learning performs better in maximizing profits.

Additionally, Tafsiri et al. [25] studies a combinatorial double auction-based marketplace that formulates the problem in form of an integer linear programming model to maximize the profit of users and providers and minimize wasted resources. In order to decrease the complexity of the approach a heuristic is applied in resource allocation mechanism. Zhang et al. [11] uses integer programming and weighted VCG algorithms to solve the static optimal problem of the supposed auction. Their auction-based mathematical model provides truthful online auction mechanism based on user evaluation and cost by applying a fast heuristic algorithm. A multi-requirement, single-minded scenario is assumed in users' requests submission. The proposed online auction mechanism while maximizing the providers' profit specifies prices based on real-time supply and demand conditions. Regarding the interests of Cloud providers and users, the users' evaluations are mapped into the payment price algorithm. The proposed scheduling and pricing framework by Babaioff [34] for Cloud resources, aimed at increasing the efficiency of Cloud resources utilization based on the economic principles. The users' requests are scheduled regarding dynamically calculated prices. Besides, this economic resource allocation framework, named ERA, provides a key internal API to modules with scheduling, pricing and demand prediction algorithms.

## 3. SYSTEM MODEL AND PROBLEM STATEMENT

Cloud computing provides a wide range of services to users all over the world; it can be considered as a marketplace with different parties including providers and users (Fig.1). In Fig.1, *Cloud Committee* has the role of the market manager consisting of two main module named *Request Dispatcher* and *Market Manager*. SaaS providers can register in the market to receive application requests through the considered API of *Request Dispatcher*. Let $Req_r$ as a request sent to the services provider. If provider $i$ can provide request $Req_r$ then it will compute its price based on the pricing policies and send the offered price called bid ($A_i$) to *Market Manager*. After that, *Cloud Committee* informs the providers the notification of the winner ($Rep_r$) who has the best offer to provide $Req_r$. Finally, based on the received notification of *Cloud Committee,* the providers update their pricing strategies to have a better bid in the future.

Requests are stored in a vector named $REQ = <Req_1, Req_2, …>$; let $Req_r$ the $r^{th}$ request in REQ. *Request Dispatcher* sends requests to available SaaS providers registered in *Cloud Committee*. It is worth mentioning that each SaaS provider is basically independent and acts as stand-alone. Since they have different computing platforms for processing the requests they might select a variety of VMs to run an application. Thus, some of the parameters are set by each SaaS provider independently, such as prices.

**Applications model:** The list of applications that provider $i$ owns is represented by $App_i = <App_{i1}, App_{i2},…, App_{iLi}>$. Each of these applications has a benefit to the provider. Thus, the list of per unit benefit of applications of provider $i$ is stored in $\beta_i = <\beta_{i1}, \beta_{i2},..., \beta_{iLi}>$, where $\beta_{ij}$ is the per unit benefit of $App_{ij}$ (application $j$ in SaaS provider $i$). $App_{ij}$ is represented in form of $<AppID_{ij}, \mu_{ij}, Srv_{ij}, \theta_{ij}, MT_{ij}>$, where $AppID_{ij}$ denotes the identification of $App_{ij}$; the application consists of a number of services represented by $\mu_{ij}$; the list of these $\mu_{ij}$ services is stored in $Srv_{ij} = \langle VMM_{j1}, VMM_{j2}, ..., VMM_{j\mu_{ij}} \rangle$. $\theta_{ij}$ is the initial price of $App_{ij}$ which is determined by the developer of the application; $MT_{ij}$ denotes the number of tenants available for the application; thus, zero is assumed for a not multi-tenant application. As mentioned previously, each SaaS provider has a set of heterogeneous VMs of different types. These VMs have a per unit benefit for the provider; the per unit benefits of VMs of provider $i$ is represented by $\alpha_i=<\alpha_{i1}, \alpha_{i2},…, \alpha_{iRi}>$ [14].

**User requests model:** The considered requests in our Cloud market are in form of applications. Therefore, the model of applications is firstly described. SaaS providers host some applications on some infrastructural resources; these resources are widely supplied by IaaS providers. Each application includes a set of services that are essential for the application to be run. In other words, these services are well defined components of functional behavior that provides a logical grouping of application functions, such as load balancing, application performance monitoring, application acceleration, auto-scaling, micro‐segmentation, service proxy and service discovery and so forth. Thus, a request ($Req_r$) is supposed as $<AppID_r, \tau_r, W_r, Srv_r>$, where $AppID_r$ is the identifier of the application; $\tau_r$ is the duration of time for $Req_r$, $W_r$ is the user's willingness to pay for this request and $Srv_r$ depicts the required infrastructural configuration of application for being run based on its services. The required infrastructure is represented by a vector of VM models (*VMM*) named $Srv_r = <VMM_1, VMM_2, …, VMM_{\mu r}>$, each *VMM* denotes the requirements of a service in the requested application; $\mu_r$ shows the number of services in $Req_r$.

**Virtual machines model:** the VM model is represented by $VMM = <Type, Core, Memory, Storage, HourCost>$, where *Type*, denotes the class of the VM as a predefined fixed label including 'small', 'medium' or 'large'; *Core* is the number of processors of the VM; *Memory* and *Storage* show the size of the RAM and the storage of the VM, respectively; *HourCost* denotes the cost of the VM per hour. For instance, a *VMM* in $Srv_r$ can be <'small', 1, 5MB, 1GB, $0.5>.

**Payment pricing:** As mentioned previously, IaaS providers have to provide required infrastructures for hosted applications on SaaS providers. In other words, IaaS providers offer VM instances for executing applications [13] and SaaS providers prepare the required infrastructures in form of VMs to host the applications. Undoubtedly, users prefer offers that have lower prices with satisfying QoS. As users cannot perceive the absolute value of an application they select the best one among the offers of SaaS providers by comparison. Resource appropriation strategies and pricing policies are of the most effective decisions of SaaS providers [35, 36] which have vital impacts on their profits.

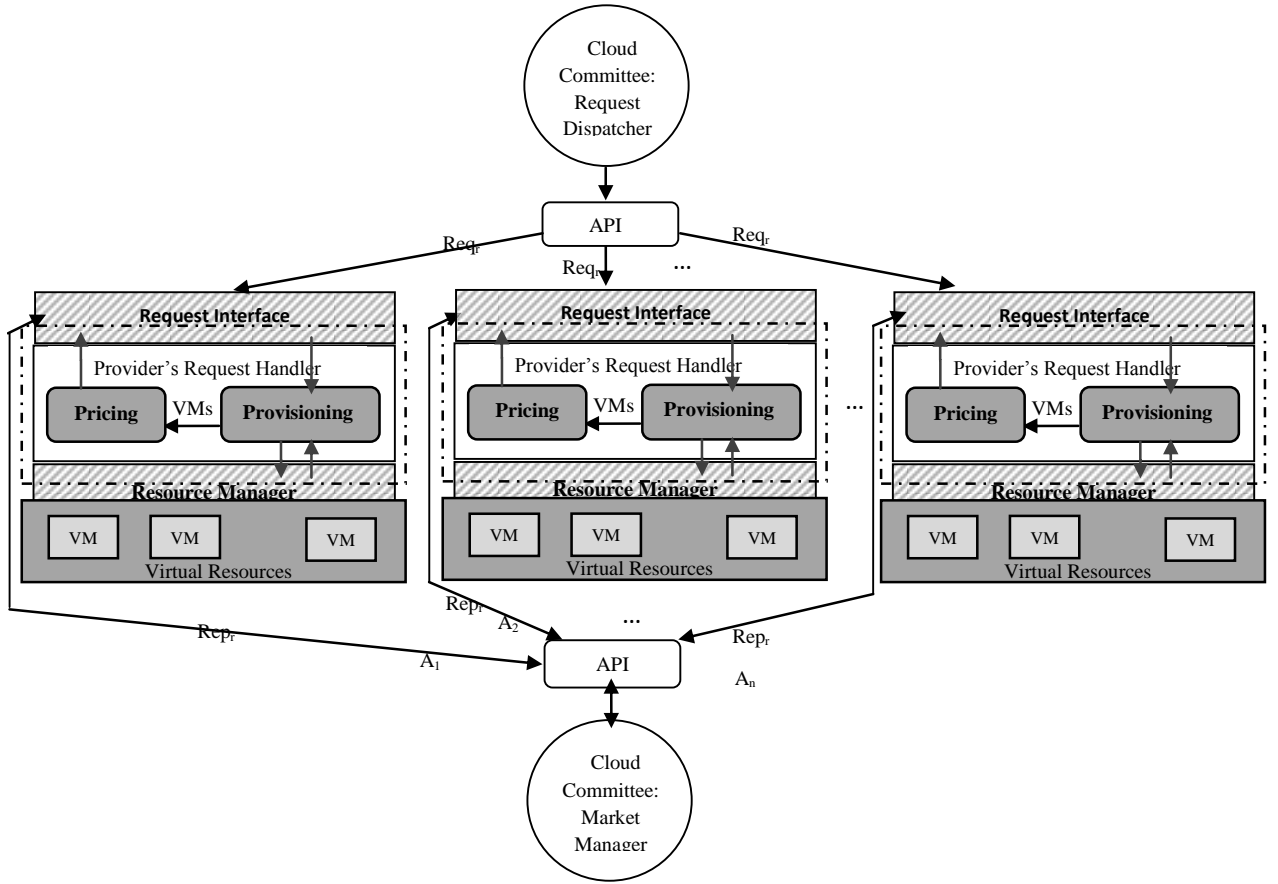

**Fig. 1** The model of Cloud marketplace

The offered price of SaaS providers basically depends on both development and deployment expenses of applications. The development expenses are license costs of the application; needless to mention, multitenant applications have higher prices, as one license can be leased to several users simultaneously. The deployment expenses are the infrastructural requirements of the applications.

Based on the aforementioned assumptions, the deployment cost of a request ($Req_r$), which demands for $App_{ij}$, is computed as

$$c_{ij} = \tau_r \times \sum_{k=1}^{\mu_{ij}} VMM_{jk}.HourCost \qquad c_{ij} \ll W_r, \forall k \in [1, \ldots, \mu_{ij}]. \tag{1}$$

The deployment cost depends on two main factors; firstly, the duration of time that $Req_r$ is run ($\tau_r$) and secondly, the costs of applied infrastructures to host the $Req_r$. The infrastructural cost of $Req_r$ is the summation of costs of all virtual machines required to run the services of the requested application per hour ($\sum_{k=1}^{\mu_{ij}} VMM_{jk}.HourCost$). As a tangible example, consider the following scenario. Suppose a typical SaaS provider has an application as <'A021', 3, <'t2.small', 'm3.large', 't2.medium'>, $340, 2>. Based on the provided prices of VM models depicted in Table2, the cost of deploying the request related to the application with identification 'A021' for $\tau$=10 hours is computed as $10Hour \times \sum_{k=1}^{3} VMM_k.hourCost = 10Hour \times \left(\frac{\$0.026}{Hour} + \frac{\$0.140}{Hour} + \frac{\$0.052}{Hour}\right) = \$2.18$.

SaaS provider *i* sends its offered price to *Market Manager* in form of $A_i$ (as depicted in Fig.1); *Cloud Committee* sends back the offered price of the winner provider to the SaaS provider *i* as $Rep_i$.

For the sake of clarity, the adopted notations in this research are summarized in Table1.

**The economics of price discrimination of applications:** In this research, the following parameters are considered in pricing policies of the SaaS providers.

· Initial cost,
· Resource appropriation,
· Multi-tenancy,

TABLE 1
SYSTEM PARAMETERS

| parameter | Declaration |
|---|---|
| $Req_r$ | Request $r$ of the Cloud market |
| $APPID_r$ | Identification number of Application $r$ |
| $A_i$ | Offered price of provider $i$ |
| $Rep_r$ | Notification of the winner of request $r$ sent by *Cloud Committee* |
| **$APP_i$** | List of application instances in provider $i$ |
| $App_{ij}$ | Properties of application $j$ in provider $i$ |
| REQ | Vector of requests |
| VMM | Model of a virtual machine |
| $L_i$ | Number of provided applications in provider $i$ |
| **$β_i$** | Per unit benefit applications of provider $i$ |
| $β_{ij}$ | per unit benefit of application $j$ for provider $i$ |
| **$α_i$** | Per unit benefit of virtual resources of provider $i$ |
| $α_{ij}$ | Per unit benefit of virtual resource $j$ for provider $i$ |
| $Srv_r$ | Set of services required for running request $Req_r$ |
| $μ_r$ | Length of $Srv_r$ or number of services required for running request $Req_r$ |
| $μ_{ij}$ | Number of services in $APP_{ij}$ |
| $u_i$ | Profit function of SaaS provider $i$ |
| $P_i$ | Offered price of SaaS provider $i$ |
| $C_i$ | Cost of providing the request by provider $i$ |
| $v_r$ | Cost of required resources for hosting request $Req_r$ |
| $v_i^t$ | Investment of provider $i$ at time $t$ |
| $ΔV$ | Value of return on investment |
| $θ_r$ | Cost of requested application specified in request $Req_r$ |
| $θ_{ij}$ | Cost of $APP_{ij}$ |
| $MT_{ij}$ | Number of users of $APP_{ij}$ (denotes multi-tenant applications) |
| $R^i_T$ | Average regret of provider $i$ in $T$ |
| $R_{max}(s_i')$ | Maximum regret that provider $i$ may experience for not playing $s_i'$ |
| N | List of providers registered in *Cloud Committee* |
| n | Number of providers registered in *Cloud Committee* |
| S | Possible strategies of players |
| $S_i$ | Possible strategies of player $i$ |
| **s** | Strategy profile of players |
| **u** | Profit function of players |
| $u_i(s)$ | Profit function of player $i$ when **s** is played |
| $s_i$ | Current strategy of player $i$ |
| $s_{-i}$ | Current strategy of all players except $i$ |
| $γ$ | Constant positive value determined by *Cloud Committee* |
| $ω_i$ | Pricing coefficient of request $Req_r$ in provider $i$ |
| q | Number of received requests of applications |
| $Ψ(S)$ | Probability distribution of strategies of players |
| $c_{ij}$ | deployment expenses of $App_{ij}$ (price of supplied VMs) |
| $W_r$ | User's willingness to pay price for $Req_r$ |
| $P^t_i(x)$ | Probability of strategy $x$ performed by provider $i$ at time $t$ |
| $τ_r$ | Duration of time $Req_r$ is required |
| T | Duration of the game |

- User's willingness to pay,
- Performance,
- Structure of payment flow.

Providers determine the price of a service based on the above mentioned parameters. *Initial price*, introducing the development expenses determined by application developers, is one of the primitive parameters involved in pricing strategies. *Resource appropriation* denotes the deployment expenses of a service; it is the second basis parameter which providers use in their pricing strategies. *Multi-tenancy*, as a single instance of a multitenant application shares development and deployment expenses while serving multiple users simultaneously a multitenant application can serve a predefined number of users; $MT_{ij}$ represents the number of users of a multitenant application. *Willingness to*

*pay* is a price determined by users while requesting, $W_r$; they expect to pay for the request less than this value [27]. If users demand for a certain level of *performance*, then providers can determine an added price for guarantee and a penalty for the violation (known as SLA). Finally, the *structure of payment flow* can affect the price as well. Payment flow can be single or recurring; providers assign fewer prices for single payments in comparison with the recurring one, which the payment is structured in multiple steps.

For more details about dynamic pricing parameters, readers are referring to [18, 20, 21].

### 3.1 Game Theory

Game theory is a powerful tool to face scenarios, including prominent interactions of rational entities. These interactions cannot be handled by classical optimization theories [5], as the behavior of a player affects not only itself but also the other players. According to Shoham [30], games include:

- $N = \{1, 2, ..., n\}$ as a finite set of players;
- $S = S_1 \times ... \times S_n$, where $S_i$ is a finite set of available strategies of player $i$;
- $\mathbf{u} = (u_1, ..., u_n)$, where $u_i(s): S \rightarrow \mathbb{R}$ is a real-valued profit function of player $i$.

Each vector $\mathbf{s} = (s_1, ..., s_n) \in S$ is called a strategy profile; thus, $u_i(\mathbf{s})$ as the profit of player $i$, not only depends on its choices ($s_i$), but also depends on the other players' choices, represented by $\mathbf{s}_{-i}$, as well. The overall goal is to find a suitable strategy profile for all players, known as a solution concept or an equilibrium. The equilibrium occurs when no player tends to change its strategy unilaterally [30, 31].

A complete information game consists of strategic interactions, where players are aware of each other's strategies and profit functions; such knowledge facilitates the process of seeking a solution. On the other hand, when a game is incomplete information, it demands stochastic techniques or learning approaches to find a solution concept [30, 31], such as regret minimization algorithm. Regret minimization algorithm is an online learning concept [37] which can achieve the solution concept while information is limited. Using this technique in a repeated game causes iteratively ignoring strategies which do not achieve a solution concept [38].

In this research, players are the providers registering in the *Cloud Committee* identified by 1, 2, ..., n. Needless to mention that the goal of registering in the committee is to reach the best possible bid in order to attract as much customer as available. Thus, the focused strategies of the registered providers are their pricing policies, i.e. $S_i$ depicts the pricing strategies of player *i*. Ultimately, to evaluate the performance of providers in choosing a particular strategy in the light of strategies of other players, profit function is used. In other words, $u_i$ gets the strategy profile of players including $S_i$ and produces a real value as the profit of player *i*. Details of the assumed game model including the strategies of players and their profit functions are discussed in section4.

### 3.2 Regret Minimization Algorithm Principles

Regret minimization has triggered a family of powerful learning algorithms; it measures regret of players after using a strategy in each round. The algorithm is a way to capture the intuition that a player performs well, no matter what others do [37, 39]. It tries to perform in a way to converge to the least value of regret for not choosing other strategies. An adaptive procedure of the algorithm sends some signals as recommended strategies to the players to compute the regrets; players compare their current profit with that of the received recommended strategy. Then, the probabilities of the strategies are updated regarding the regrets. Players are expected to achieve a strategy distribution called correlated equilibrium, known as the most relevant non-cooperative solution concept [30], as follows.

*Definition1* Probability distribution $\Psi$ on $S$ is a correlated equilibrium of $G$ if, for every player $i \in N$, and every strategy profile $s$, $s_i' \in S_i$, we have [40]:

$$\sum_{s \in S} \Psi(S)[u_i(s_i', \mathbf{s}_{-i}) - u_i(\mathbf{s})] \leq 0, \tag{2}$$

where $\psi(S)$ denotes the distribution of strategies of players represented as $S$; $u_i(s_i', \mathbf{s}_{-i})$ is the profit of player *i* for choosing strategy $s_i'$ when other players play $\mathbf{s}_{-i}$; $u_i(\mathbf{s})$ denotes the profit of player *i* when strategy profile $\mathbf{s}$ is chosen; as mentioned previously $\mathbf{s} = (s_1, ..., s_n)$. The inequality of Eq.2 denotes that player *i* does not obtain a higher profit by choosing strategy $s_i'$ instead of its current strategy $s_i$.

Every finite game has a nonempty, closed and convex set of correlated equilibrium [30]; according to Proposition1, a proper adaptive procedure ensures that the distributions of the game will converge to the set of correlated equilibrium with probability one [41].

**Proposition1** If a game runs iteratively many times and each player performs according to a certain regret minimization strategy, then the empirical distribution of the game will converge to a set of correlated equilibrium.

**Corollary** The proof is based on a recursive formula for the distance of the vector of regrets to the negative orthant. Many iteration of a game does not mean a simple one-period recursion from $t$ to $t+1$. Instead a multi-period recursion where a large "block" of periods, from $t$ to $t+v$, is combined together. Both $t$ and $v$ are carefully chosen and go to infinity; but, $v$ is relatively small in comparison with $t$. Consider $\prod_t$ as the probabilities of transitioning from $t$ to $t+1$.

$$\prod_t(j,k) = \begin{cases} \dfrac{1}{\mu} R_t(j,k), & \text{if } k \neq j, \\ 1 - \sum_{k' \neq j} \dfrac{1}{\mu} R_t(j,k'), & \text{if } k = j. \end{cases}$$

Note that $\mu$ guarantees that the probabilities have greater values than zero for all possible values of $j$ and $t$.

Let $m$ as the number of strategies of player $i$ and $M$ as the upper bound of payoffs of player $i$. Let $L = \{(j,k) \in S^i \times S^i : j \neq k\}$; therefore, $\mathbb{R}^L$ is the Euclidean space with $m(m-1)$ dimensions. Then, for $t>0$ we have the following.

$$A_t(j,k) = 1_{\{s_t^i = j\}} [u^i(k, s_t^{-i}) - u^i(s_t)],^1$$
$$D_t(j,k) = \frac{1}{t} \sum_{\tau=1}^{t} A_\tau(j,k),$$
$$R_t(j,k) = D_t^+(j,k) \equiv [D_t(j,k)]^+.$$

Finally, let $\rho_t = [dist(D_t, \mathbb{R}_-^L)]^2$. Since the closest point to $D_t$ in $\mathbb{R}_-^L$ is$^2$ $D_t^-$, $\rho_t = \|D_t - D_t^-\|^2 = \|D_t^+\|^2 = \sum_{j \neq k} [D_t^+(j,k)]^2$.

The proof of proposition1 has 11 steps as follows.

**Step 1:**

(i) $\quad E[(t+v)^2 \rho_{t+v} | h_t] \leq t^2 \rho_t + 2t \sum_{w=1}^{v} R_t \cdot E[A_{t+w} | h_t] + O(v^2);$

it is the basic recursion equation.

(ii) $\quad (t+v)^2 \rho_{t+v} - t^2 \rho_t = O(tv + v^2).$

Define

$$\alpha_{t,w}(j, s^{-i}) = \sum_{k \in S^i} \prod_t(k,j) P[s_{t+w} = (k, s^{-i}) | h_t] - P[s_{t+w} = (j, s^{-i}) | h_t].$$

**Step 2:**

$$R_t \cdot E[A_{t+w} | h_t] = \mu \sum_{s^{-i} \in S^{-i}} \sum_{j \in S^i} \alpha_{t,w}(j, s^{-i}) u^i(j, s^{-i}).$$

The coefficients $\alpha$ depend on the moves of the other players. Indeed, given $h_t$, the choices $s_{t+w}^i$ and $s_{t+w}^{-i}$ are not independent when $w>1$ (since the transition probabilities change with time). Therefore, we replace the process $(S_{t+w})_{0 \leq w \leq v}$ by another process $(\hat{S}_{t+w})_{0 \leq w \leq v}$, with a stationary transition matrix (that of period $t$).

**Step 3:**

$$R_{t+v}(j,k) - R_t(j,k) = O\left(\frac{v}{t}\right).$$

For $t>0$ and $h_t$, an auxiliary stochastic process $(\hat{S}_{t+w})_{w=0,1,\ldots}$ with values in $S$ as follows:

The initial value is $\hat{S}_t = S_t$, and the transition probabilities are $P[\hat{S}_{t+w} = s | \hat{S}_t, \ldots, \hat{S}_{t+w-1}] = \prod_{i' \in N} \prod_t^{i'} (\hat{S}_{t+w-1}^{i'}, s^{i'}).$

The $\hat{s}$-process is thus stationary: It uses the transition probabilities of period $t$ at each period $t+w$, for all $w \geq 0$.

**Step 4:**

$$P[s_{t+w} = s | h_t] - P[\hat{s}_{t+w} = s | h_t] = O\left(\frac{w^2}{t}\right).$$

Define

$$\hat{\alpha}_{t,w}(j, s^{-i}) = \sum_{k \in S^i} \prod_t(k,j) P[\hat{s}_{t+w} = (k, s^{-i}) | h_t] - P[\hat{s}_{t+w} = (j, s^{-i}) | h_t].$$

For $w$ small relative to $t$, the change in probabilities is small (see the Steps 3 and 4).

**Step 5:**

---

$^1$ $1_G$ is the indicator of the event $G$.
$^2$ $[x]^-$ means $\min\{x,0\}$.

$$\alpha_{t,w}(j,s^{-i}) - \hat{\alpha}_{t,w}(j,s^{-i}) = O\left(\frac{w^2}{t}\right).$$

The total difference is estimated.

**Step 6:**

$$\hat{\alpha}_{t,w}(j,s^{-i}) = P\left[s_{t+w}^{\hat{}-i} = s^{-i}|h_t\right][\Pi_t^{w+1} - \Pi_t^w](s_t^i,j),$$

where $\Pi_t^w \equiv (\Pi_t)^w$ is the $w^{th}$ power of the matrix $\Pi_t$, and $[\Pi_t^{w+1} - \Pi_t^w](s_t^i,j)$ denotes the $(s_t^i,j)$ element of the matrix $\Pi_t^{w+1} - \Pi_t^w$.

The moves of the other players (which, in the $\hat{s}$-process, are independent of the moves of player $i$) are factored out from the coefficients $\hat{\alpha}$. At this point we get the difference between the transition probabilities after $w$ periods and after $w+1$ periods.

**Step 7:**

$$\hat{\alpha}_{t,w}(j,s^{-i}) = O(w^{-1/2}).$$

The difference depicted in Step 6 is shown to be small, since $w$ is large and the transition matrix has all its diagonal elements strictly positive.

**Step 8:**

$$E[(t+v)^2 \rho_{t+v}|h_t] \leq t^2 \rho_t + O(v^3 + tv^{1/2}).$$

For each $n=1,2,\ldots$, let $t_n = \lfloor n^{5/3} \rfloor$ be the largest integer not exceeding $n^{5/3}$. Substituting in (i) in Step 1 yields the final recursive formula.

**Step 9:**

$$E[t_{n+1}^2 \rho_{t_{n+1}}|h_t] \leq t_n^2 \rho_{t_n} + O(n^2).$$

**Step 10:**

$$\lim_{n \to \infty} \rho_{t_n} = 0 \quad a.s.$$

**Step 11:**

$$\lim_{t \to \infty} R_t(j,k) = 0 \quad a.s.$$

If a carefully chosen subsequence of periods $(t_n)_{n=1,2,\ldots}$ is considered the proof ends. It is worth mentioning that the proof of described steps can be found in [41]. □

Based on Proposition 1, the empirical distributions converge to the set of correlated equilibrium if and only if the average of regrets converges to zero. The regret of player $i$ for strategy profile $s$ is computed using Eq.3, as follows.

$$\left[\max_{s_i' \in S_i} u_i(s_i', s_{-i})\right] - u_i(s). \qquad (3)$$

The regret is the profit that player $i$ has lost by playing $s_i$ instead of playing the recommended strategy $s_i'$. In other words, the difference of profit of player $i$ for playing $s_i'$ instead of $s_i$ is assumed as its regret. To clarify this equation, consider a *Cloud Committee* with 5 SaaS providers; let $u_1(s) = \$100$ as the strategy of player 1, and the recommended strategy that has the greatest profit is $[\max_{s_1' \in S_1} u_1(s_1', s_{-1})] = \$93$; other players' strategies are \$107, \$96, \$110, and \$103. The regret of player 1 is $u_1(\$93, \$107, \$96, \$110, \$103) - u_1(\$100, \$107, \$96, \$110, \$103)$.

## 4. OPTIMIZATION PROBLEM FORMULATION AND SOLUTION

In this section, firstly, the formulation of SaaS providers for optimizing their profit is presented. Then, the optimization problem is studied in form of an incomplete information game, and the equilibrium properties are discussed by the aim of regret minimization techniques. Finally, the economic impact of the considered game is discussed.

### 4.1 Providers' Profits Optimization Formulation

As previously mentioned, SaaS providers compete with each other for users' requests, named $Req_r$; the requests are received from *Request Dispatcher*. In such conflicting situation, the providers face an optimization issue of profit maximization while keeping users satisfied as well. Provider $i$ computes the costs of providing the request, and it offers a price, $A_i$, to *Market Manager* (Fig.1). The *Market Manager* evaluates the received offers of all registered providers and notifies the winner to all providers, named $Rep_r$. Finally, an agreement is formed between the winner provider and the user who has sent $Req_r$.

SaaS providers compete with each other to attract as more users as possible; this is achieved by offering a desired price. The offered price of a provider depends on previous mentioned criteria, which dynamically determine the price. The profit of SaaS providers is the surplus remaining after total costs are deducted from total revenue. The total costs associated with a SaaS provider ($C_i$) include the potential costs of software licenses, providing infrastructures, service disruptions, data security concerns, etc. [19, 23]; the total revenue that a SaaS provider obtains is determined on the basis of the pricing policies of the provider $i$ depicted by $P_i$. The profit of provider $i$ ($u_i$) is formulated as

$$u_i = P_i - C_i, \tag{4}$$

where, $P_i$ and $C_i$ are the revenue of SaaS provider $i$ and the associated cost of software licenses, infrastructures and so forth, respectively. Providers have to consider some constraints while determining the prices. The considered constraints of SaaS provider $i$ are formulated in form of an optimization problem as follows.

$$\begin{aligned} max\ u_i &= max(P_i - C_i) = max\ P_i - min\ C_i \\ s.t\quad &P_i \leq W_r, \\ &P_i \leq \theta_{ij} + c_{ij}, \\ &P_i > 0, C_i > 0. \end{aligned} \tag{5}$$

The constraints are settled to guarantee some features as follows. The first constraint ($P_i \leq W_r$): the offered price ($P_i$) should not exceed the user's willingness to pay ($W_r$ introduced as a parameter in $Req_r$) [42], otherwise demands for the application will decrease as users stop buying that. It is worth noticing that user's willingness to pay mostly exceeds the initial price of the application and its deploying cost [27]. Second constraint ($P_i \leq \theta_{ij} + c_{ij}$): the offered price cannot exceed the sum of $\theta_{ij}$ and $c_{ij}$, where $\theta_{ij}$ introduces the price of supplied VMs for deployment of the requested application $App_{ij}$ showed in Eq.1, and $c_{ij}$ is the price of the requested application $App_{ij}$, respectively. If this constraint is violated, then the user may prefer to buy the application from its developer and deploy it on some offered resources from IaaS providers, instead of demanding from SaaS providers. The last constraint ($P_i > 0$ and $C_i > 0$): the price and the cost, both have positive values.

Each provider tries to offer the lowest price to become the winner in *Cloud Committee*. Thus, they offer prices lower and lower until they reach zero. It is logical to assume such an outcome; however due to these constraints, the game does not converge to zero-cost applications.

*4.2 Introduction to the Game Model*

Interaction of providers is known as a significant challenge in Cloud computing marketplace, especially when complete information is not available [7]; the game theory has successfully been applied to solve such issue. In this research, we overcome the problem of conflicting situation with a game-theoretic approach. The approach analyzes and optimizes pricing strategies of applications. Definition2 introduces the considered game of this study as follows.

*Definition2:* Let $G = (N, S, u)$ as the non-cooperative game introduced for the application pricing problem. $N$ is a finite set of $n$ SaaS providers that enroll *Cloud Committee*; they are indexed by $i$; $S=S_1\times\cdots\times S_n$, where $S_i$ is the strategy set of player $i$; $\boldsymbol{u} = (u_1, ..., u_n)$ where $u_i(s)$: $S\rightarrow\mathbb{R}$ is known as the profit function of player $i$ who selects its strategy $s_i \in S_i$, in a way to maximize its profit; vector $\boldsymbol{s}$, as the strategy profile of the game, consists of the selected pricing strategies of all providers.

SaaS providers have no information about the strategies of other providers; they are only informed of the winner. To overcome such incomplete information game, it has been equipped with a learning model, which specifies the learning rules of the providers and affects their interactions. Regarding the non-stationary environment, an adaptive decision making technique, named regret minimization technique, is applied. This technique is a suitable learning algorithm in such scenario and it guarantees convergence to an equilibrium in incomplete information games (Proposition1).

*4.2.1 Profit function in the Game*

As mentioned previously, a SaaS provider with the minimum offered price for the requested application is the current winner of the game. Since the trade-off between achieving high revenues and spending low costs is an important issue, both parts must be taken into the profit function as follows.

*Definition3:* For the strategy profile $s \in S$, $u_i(s)$: $S\rightarrow\mathbb{R}$ is the profit function, which assigns numerical values to strategy profile $S$. For all strategy profiles $\boldsymbol{x}, \boldsymbol{y} \in S$, provider $i$ prefers $\boldsymbol{x}$ over $\boldsymbol{y}$ iff $u_i(\boldsymbol{x}) > u_i(\boldsymbol{y})$. It is to be noted that,

$u_i(x)$ not only depends on strategy of provider $i$ but also relates to all strategies of other providers; needless to mention that strategy profile $x = (x_1, \ldots, x_i, \ldots, x_n)$. Thus, the profit function of SaaS provider $i$ is

$$u_i = \begin{cases} S_i - C_i, & argmin(s) = i \\ 0, & else \end{cases}, \tag{6}$$

where, $S_i$ denotes the pricing strategy of provider $i$ for $Req_r$, written as $P_i$ in Eq.4; as previously mentioned, $Req_r$ is demanding for $App_{ij}$. The profits of all players except the winner of the game are zero; the winner is a player whose offered price is the least. $S_i$ is determined as the following equation.

$$S_i = \sqrt{\omega_i}(1 + \gamma\sqrt{\omega_i})(\theta_{ij} + c_{ij}), \tag{7}$$

where, $\omega_i$ is a parameter which is specified by provider $i$, $\gamma$ is a constant positive value determined by the *Cloud Committee*; it is less than 1. $\theta_{ij}$ and $c_{ij}$ are the price of the requested application $App_{ij}$ and the price of supplied VMs for deployment of the requested application $App_{ij}$, respectively. Players choose their strategies by choosing different values of $\omega_i$ in a way to optimize their profits. The explanation of how provider $i$ opts for $\omega_i$ is provided in section 5.1 Experimental setup.

$C_i$ in Eq.6 denotes the cost of serving the request $Req_r$ demanding $App_{ij}$ as

$$C_i = \alpha_{ij} c_{ij} + \beta_{ij} \theta_{ij}. \tag{8}$$

$\alpha_{ij} \in \alpha_i$ is per unit benefit of application $App_{ij}$ hosted by provider $i$ and $\beta_{ij} \in \beta_i$ is per unit benefit of virtual resources that provider $i$ has applied for deploying application $App_{ij}$. Offering each application and each virtual machine has a benefit for the SaaS provider stored in $\alpha_i$ and $\beta_i$ sets, in the order mentioned.

Eventually, by substituting Eq.7 and Eq.8 in Eq.6, $u_i$ is formally expanded as

$$u_i = \begin{cases} \sqrt{\omega_i}(1 + \gamma\sqrt{\omega_i})(\theta_{ij} + c_{ij}) - (\alpha_{ij} c_{ij} + \beta_{ij} \theta_{ij}), & argmin(s) = i \\ 0, & else \end{cases}. \tag{9}$$

Algorithm1 presents the considered game between SaaS providers. The algorithm is run in a distributed manner by each SaaS provider in Cloud market separately as providers are mainly dependent; the provider gets the request of a user as its input ($Req_r$) and the output is sent to *Market Manager* in form of *BidList*. Then, in a loop that ends when the equilibrium is achieved the following steps are run. Firstly, *VMsSelection* procedure gets the request ($Req_r$) and returns a list of suitable virtual machines named *SelectedVMList* that map the requirements of $Req_r$. In the next step, this provider must offer price for providing the requirements of $Req_r$ shown in *BidList*. The provider runs the procedure called Bid according to the $Req_r$ and the list of the selected virtual machines (*SelectedVMList*). The provider computes its profit using *Profit* procedure based on the offered bid (*ProfitList*). After that in line 4, the *BidList* is sent to the *Market Manager* as depicted in Fig.1 by $A_i$. Then in line 5, the *Market Manager* finds the *Winner* provider whose bid is the best based on the price and its accommodation to the requirements by running *MarketMgr* procedure; in the case that a provider is the *Winner* then it should allocate required resources to $Req_r$ (line 6). Whenever the equilibrium circumstances occur Algorithm1 terminates in line 9. During this loop, each provider tries to offer a bid which leads to making the provider as the *Winner*.

Now we present our theoretical results on time complexity of Algorithm1 in lemma1.

**Lemma1** The time complexity of Algorithm1 is $O(|S|)$ at round $t$ for provider $i$, where $|S|$ is the number of pricing strategies of the provider $i$. More specifically, if the underlying game satisfies the equilibrium condition, then its time complexity in processing each request is $O(|S|^2 n)$ on average, where $n$ is the number of providers.

**Proof** Based on Proposition1, this algorithm will converge to the set of correlated equilibrium. $|S|$ is the number of strategies player $i$ has; each of the regrets is bounded by an expression that depends on the maximum regret that provider $i$ may experience for not playing a recommended strategy. In order to minimize the average regret of players, a recommended strategy is chosen from $|S|$ strategies. Although the average regret of players changes in every step, it is bounded by the upper bound, the algorithm must terminate. Since the least difference between any two regrets is bounded, the result of the lemma follows directly. Note that Lemma1 also states that having knowledge about other providers' strategies makes the algorithm better informed and decreases its run-time. As the game is running among *n* players, thus the time complexity of the algorithm is $O(|S|^2 n)$, an upper bound on the complexity of Algorithm1. □

**ALGORITHM1** RPG: PRICING OPTIMIZATION GAME ALGORITHM

The algorithm is run by each SaaS provider joined in the *Cloud Committee*, as CurrentPrv, in a distributed manner.

    **Input:**
        *Req$_r$*
    **Output:**
        BidList
    **Do**
1.     SelectedVMList = VMsSelection(*Req$_r$*);
2.     BidList[CurrentPrv] = Bid(*Req$_r$*, SelectedVMList);
3.     ProfitList[CurrentPrv] = Profit(CurrentPrv, BidList);
4.     Send BidList[CurrentPrv] to Market Manager;
5.     Winner = MarketMgr(BidList);
6.     **If** currentPrv **matches** Winner **then**
7.         Allocate(Req$_r$, SelectedVMList);
8.     **End**
9. **until** Equilibrium is reached

## 4.3 Theory on Market Equilibrium

In this Section, we study the properties of the solution concepts of the considered pricing game. Firstly, the conditions on the equilibrium are discussed; then, the usage of regret minimization algorithm is studied.

### 4.3.1 Proposed Regret Minimization Algorithm

In an incomplete-information game, players have correlated preferences; thus, a coordinator is required to help the players select the strategies in a way to reach the equilibrium [40]. The coordinator suggests strategies based on regret minimization algorithm (Section3.3.1). In such a case, regret minimization algorithms can be applied to enable players to acquire some knowledge of the game. Regret minimization algorithms have some benefits, such as no need to coordinate between the players, or no necessity for the players to know the others' profits [40]. At each round of the game, players may play the same strategy as in the previous period, or they may switch to other strategies with some probabilities that are commensurate with their profits [40]. Let $U$ be the total profit up to now, and $U(s')$ be the total profit, obtained if $s'$ was played instead of the current strategy. If $U(s') > U$ then, strategy $s'$ would be better to exchange with current strategy; the probabilities of the strategies are updated based on the regret of player for not playing $s'$. We propose some new equations for updating the probabilities of strategies in a way to reach no regrets in a shorter time [39]. Furthermore, this work studies different regret minimization techniques including external, internal, and swap.

As previously mentioned, in order to reach equilibrium in the game, the providers update the probabilities of their strategies regarding their average regrets in the hope that they will experience less regrets in the future. Thus, the probabilities are improved based on the average regrets experienced in the duration of time (*T*). The average regret of provider *i* in *T*, for playing $s_i$, when the coordinator recommends $s_i'$ is

$$R_T^i(s_i, s_i') = max\left\{\frac{1}{T}\sum_{t \leq T}\left[max_{s_i' \in S_i} u_i(s_i', \mathbf{s}_{-i}) - u_i(\mathbf{s})\right], 0\right\}. \tag{10}$$

Regarding Eq.3, regret of provider *i* for strategy profile $s$ is $max_{s_i' \in S_i} u_i(s_i', \mathbf{s}_{-i}) - u_i(\mathbf{s})$, where $u_i(s_i', \mathbf{s}_{-i})$ is the profit of provider *i* if it plays $s_i'$ instead of $s_i$, while others' strategies remain unchanged ($\mathbf{s}_{-i}$) and $u_i(\mathbf{s})$ denotes the profit of provider *i* when strategy profile $s$ is chosen. It is to be noted that *max* is applied to avoid negative regrets. As a tangible example, consider this scenario that in *T*= [1, ..., 4], provider 1 experiences the following regrets for not playing $93 instead of $100 (using Eq.3), regrets = {$50, $0, $ 130, $74}. Then, according to Eq.10, the average regret of provider *i* during T (*t* = 1,… ,4) for not playing the strategy $93 instead of the strategy $100 is $\frac{50+0+130+74}{4}$ = $63.5. To update the probabilities of strategies of provider *i*, the following equation is introduced.

Finally, according to the obtained regret, provider *i* updates the probability of its strategies. The following equations are applied to update the probabilities of strategies of provider *i* at time *t*+1.

$$p_i^{t+1}(x) = \begin{cases} p_i^t(x)\left(1 - \frac{R_T^i(s_i, s_i')}{R_{max}(s_i')}\right) & , \forall x = s_i \\ \frac{R_T^i(s_i, s_i')}{R_{max}(s_i') \times (|S|-1)} + p_i^t(x)\left(1 - \frac{R_T^i(s_i, s_i')}{R_{max}(s_i')}\right) & , else \end{cases}, \quad (11)$$

where, $p_i^t(x)$ represents the probability of strategy $x$ at time $t$; $R_{max}(s_i')$ denotes the maximum regret that provider $i$ may experience for not playing $s_i'$; $|S|$ is the number of possible strategies that provider $i$ has. If the set of strategies of providers is infinite then term $\frac{R_T^i(s_i, s_i')}{R_{max}(s_i') \times (|S|-1)}$ will be zero, without loss of generality; the larger experienced regret for strategy $s_i$, the smaller assigned probability for the strategy $s_i$ and vice versa. The probabilities are initially assigned the same values and during the run of the game they are updated to converge to an optimal point. The applied equation in Eq.11 is derived from update equations of learning automata actions in the case that the reinforcement signal of the environment for undesirable replies [33]. In Eq.11, the penalty parameter of learning automata is replaced with a variable value based on the regret $\left(\frac{R_T^i(s_i, s_i')}{R_{max}(s_i')}\right)$. All the generated probabilities in this equation are between 0 and 1 inclusive.

As previously mentioned, regret minimization algorithm must compare the current choice in each round with a not chosen strategy to find the regret; a recommender introduces the not chosen strategy. The profit of the introduced strategy of the recommender, known as $s_i'$, is compared with the gained profit of playing current strategy, named $s_i$. Regarding the way that the recommender performs, three types of regret minimization algorithms exist including external, internal and swap regret minimization technique. In external regret, the coordinator recommends a predefined strategy to all players, without paying attention to the current strategy. The coordinator in internal regret uses a mapping between each of the strategies; i.e. for each strategy a predefined strategy is recommended. Thus, a particular strategy is recommended for the provider's current strategy and the provider finds the benefits of choosing a particular strategy in comparison with current one. Finally, the swap regret minimization coordinator randomly recommends a strategy instead of the current strategy.

### 4.4 The Economic Impact of the Considered Game

In this section, the economic impact of our pricing game is discussed. According to the reports of Eurostat [43], Cloud computing services used by one out of every five enterprises in 2013; furthermore, by 2015, revenue of SaaS is $27.94 billion, compared against $13.46 billion in 2011. Such increases in revenue demonstrate the importance of economic impact of pricing policies of Cloud providers. Based on ProfitBricks report, just 1% improvements in pricing strategies will boost the investments 11.1% in average [44]. In the light of the aforementioned facts outlined in the preceding paragraphs (Section4.1), provider $i$ has a budget constraint as $C_i \leq S_i$, which must be noted while pricing the services.

It is to be noted that billing of the services of registered users occurs periodically (usually monthly [45]). Based on the performed billing, the investment or capital of providers can be computed using Eq.12. The investment of provider $i$ is computed as

$$\begin{aligned} V_i^t &= V_i^{t-1} + \sum u_i, \\ V_i^t &= V_i^{t-1} + \sum S_i - \sum C_i, \\ V_i^t &= V_i^{t-1} + \sum \sqrt{\omega_i}(1 + \gamma\sqrt{\omega_i})(\theta_{ij} + c_{ij}) - \sum(\alpha_j c_{ij} + \beta_{ij}\theta_{ij}). \end{aligned} \quad (12)$$

$V_i^t$ represents the investment of provider $i$ at time $t$ which changes based on the current achieved profit ($u_i$). Then, according to the equation of profit shown in Eq.6, $u_i$ is substituted by $S_i - C_i$. Finally, based on the equations of $S_i$ and $C_i$ presented in Eq.7 and Eq.8 respectively, Eq.12 is rewritten.

As mentioned before, our presented pricing mechanism in this article computes rough values for undercharging and overcharging a user. For instance, the offered prices of providers do not include costs for marketing staff, or sales management. Let $q$ as the number of received requests of applications, $L_i$ as the number of provided applications in provider $i$, and $n$ as the number of providers registered in *Cloud Committee*. The provider's profit will be a positive value if it wins the game, otherwise it will be zero; so, for provider $i$, the probability of getting success in $q$ independent yes/no trials, each of which yields success with probability $\frac{1}{n}$, is given by a binomial distribution; i.e. on average, the provider would win the game with probability $\frac{1}{n} \times q$. Due to this probability, the overall provider's profit presented in Eq.9 is rewritten as $\frac{q}{n} \times E(\sqrt{\omega_i}(1 + \gamma\sqrt{\omega_i}) \times (\theta_{ij} + c_{ij}) - \sum \alpha_j c_{ij} + \beta_{ij}\theta_{ij})$, where $E(.)$ refers to the expected value. The average of $\sum \alpha_j c_{ij} + \beta_{ij}\theta_{ij}$ would be $L_i \times E(\alpha_j c_{i(1..L_i)} + \beta_{ij}\theta_{i(1..L_i)})$. Eq.12 is rewritten as

$$\overline{\Delta V} = \frac{q}{n} \times E(\theta + c) \times E\left(\sqrt{\omega}(1 + \gamma\sqrt{\omega})\right) - L_i \times E(\alpha c + \beta\theta).$$

The positive $\overline{\Delta V}$ denotes the provider has a satisfying value of return on investment. To have a positive $\overline{\Delta V}$, the following equation must be satisfied; $\frac{q}{n} \times E(\theta + c) \times E\left(\sqrt{\omega}(1 + \gamma\sqrt{\omega})\right) > L_i \times E(\beta\theta + \alpha c)$, which leads to

$$q \gg \frac{L_i \times n}{E\left(\sqrt{\omega}(1+\gamma\sqrt{\omega})\right)}. \tag{13}$$

It shows that the profit of provider $i$ improves when $q$ reaches a value greater than $\frac{L_i \times n}{E\left(\sqrt{\omega}(1+\gamma\sqrt{\omega})\right)}$. Besides, the larger the values of parameter $\omega$, the provider would get the faster return on investment; however, the amount of loss may be compensated in time.

## 5. PERFORMANCE EVALUATION

In this section, some experiments are designed to analyze the proposed model of Cloud computing marketplace and to evaluate its functionality. In our considered scenario, the winner of the competition is chosen by *Committee* instead of the user choice without loss of generality. Firstly, the parameter settings and the simulation configuration are studied. Then, the evaluation experiment of the approach is presented.

### 5.1 Experimental setup

The experimented Cloud computing marketplace is modeled by randomly generating number of SaaS providers, requests, VMs of each provider and so forth; a number from 5 to 100 to simulate number of SaaS providers with entering 1 to 100 requests per second. Although there are a growing number of SaaS providers, a bounded number is reported [1]. Initially, SaaS providers own a random number of VMs, from 100 to 1000; the types of considered VMs are based on what defined by Amazon EC2. The parameters of the VMs, previously introduced as VM model (VMM), are listed in Table2.

The processes in SaaS providers have two stages. The first one finds the most proper combination of VMs to host each service of the requested application. In this stage a VM, with a low cost per hour, which can host the service is chosen. In second stage, the provider runs the pricing strategy to offer a price for providing the request. To implement the model, a variety of changes are made in some CloudSim classes and some new ones are added to the simulator. Due to the different characteristics of virtual machines, various number of VM instances with different configurations are considered as Cloud instances, shown in Table2. The pricing game scenario simulation is implemented on Matlab 2007 on an Intel core 2 Duo CPU with 4GB RAM.

The number of applications in each provider is from 10 to 100. Our considered applications in this research are ERP (Enterprise Resource Planning), being provided by different SaaS providers. One of these ERP applications is CRM, which has three main instances: Essential, Basic, and Professional. Some instances of the potential costs of Microsoft CRM applications are presented in Table3. Costs of providers' applications varies based on the commercial fees [46]. Providers are monthly billed per user for online provisioning. Besides, for an on-premise provisioning the licensing price is determined based on the instances. Initial price of applications is determined by SaaS providers, who offer ERP, like Actionstep, iCIMS, Plex Systems and Host Analytics Inc.; the assumed values of parameters of the simulation are derived from [19, 23], regarding the commercial fees applied by IaaS providers [47], and by SaaS providers [46].

In our experiments, as γ corresponds to 0.05 interest rate it is set to 0.95 [14]. The probability distribution of $\omega_i$ is initialized as uniform distribution. It is derived from a finite set based on the level of the service.

Providers choose their strategies by choosing different values of $\omega_i$ in a way to optimize their profits. After determining the price of an application (Eq.7), provider $i$ computes its profit (Eq.9); then, it computes the regret of not choosing the recommended strategy instead of its current one (Eq.10). Eventually, the probabilities of strategies are updated (Eq.11) to experience fewer regrets in the future. The game continues until some strategies, known as equilibrium points, are found. Thus, the simulation runs until reaching these points.

TABLE 2
VMs DEFINED BY AMAZON EC2 AS AN IaaS PROVIDER (IN DECEMBER 2015) [45]

| Attr. Size | VCPU | Memory (GB) | Storage (GB) | Price per VM/$ |
|---|---|---|---|---|
| t2.small | 1 | 2 | 1x 4 SSD | $0.026/Hour |
| t2.medium | 2 | 4 | 1x 4 SSD | $0.052/Hour |
| m3.medium | 1 | 3.75 | 1x 4 SSD | $0.070/Hour |
| c3.large | 2 | 3.75 | 2x 16 SSD | $0.105/Hour |
| m3.large | 2 | 7.5 | 1x 32 SSD | $0.140/Hour |
| R3.large | 2 | 15 | 1x 32 SSD | $0.175/Hour |

TABLE 3
CONSIDERED APPLICATIONS OFFERED BY SaaS PROVIDERS WITH THEIR COSTS [36]

| Type of Provision Application's License | On-Premise | Online (per user per month) |
|---|---|---|
| CRM Server 2013 | $4922 | $150 |
| CRM Professional User CAL | $983 | $65 |
| CRM Professional Device CAL | $787 | $65 |
| CRM Basic User CAL | $342 | $30 |
| CRM Basic Device CAL | $236 | $30 |
| CRM Essential CAL | $79 | $15 |

## 5.2 Equilibrium Efficiency

The experiments are executed on a Cloud marketplace with five SaaS providers; the profits of the providers are examined while receiving 100 different requests. As the game proceeds, the profits of the providers increase as well, as depicted in Fig.2. The figure compares equilibrium efficiency of different techniques of regret minimization on pricing strategies of providers in the marketplace. Without loss of generality, the number of SaaS providers and requests are assumed very low. To show the truthfulness of our results, at the end of this section the experiments with much higher number of SaaS providers and requests are provided.

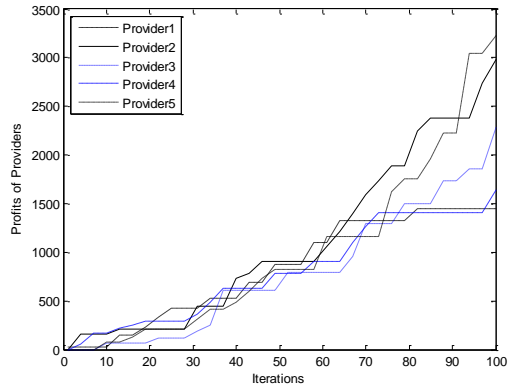

(a)

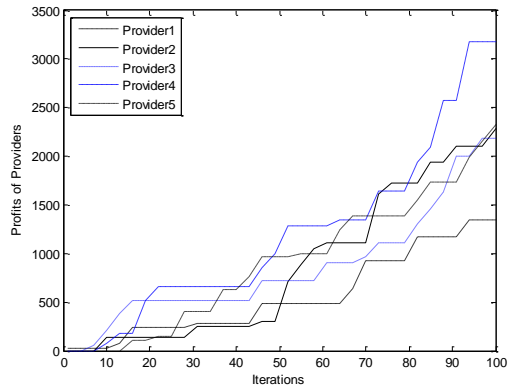

(b)

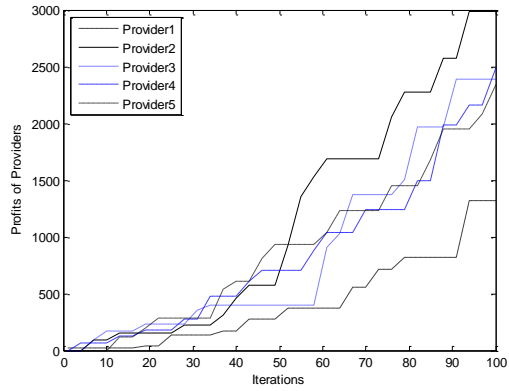

(c)

**Fig. 2** Profits of SaaS providers using **(a)** External regret minimization **(b)** Internal regret minimization **(c)** Swap regret minimization

In this research, the external regret coordinator recommends SaaS provider, a strategy which makes the offered price as small as possible. The coordinator in internal regret uses a mapping between each pair of the strategies; it recommends a particular strategy instead of the provider's current strategy; the internal recommended strategy is introduced based on permutation of number of strategies that provider $i$ has. Finally, the coordinator in swap regret minimization technique randomly recommends a strategy instead of the current strategy of the provider; i.e. there will be different pairs of strategies recommended; the external recommender introduces a strategy which makes the offered price as small as possible.

TABLE 4
CHANGES OF THE PROFITS OF PROVIDERS USING DIFFERENT TECHNIQUES

| Different Pricing Techniques | External R.M. | Internal R.M. | Swap R.M. | Non-competition-based Pricing | RM with learning function in [30] | Pricing policies in [9] | Pricing policies in [19] | Random Pricing Techniques |
|---|---|---|---|---|---|---|---|---|
| Profit in Iteration 25 | 73.98016 | 68.4607 | 67.52008 | 51.8364 | 38.97911 | 129.8065 | 68.1981 | 32.7065 |
| Profit in Iteration 50 | 135.31609 | 111.55481 | 86.11657 | 47.2032 | 69.43673 | 129.8065 | 89.736 | 36.8616 |
| Profit in Iteration 75 | 178.345291 | 153.86802 | 126.3146 | 48.6688 | 98.0287 | 129.8065 | 112.9271 | 42.3701 |
| Profit in Iteration 100 | 228.82859 | 204. 5176 | 192.1307 | 52.47 | 143.6948 | 129.8065 | 149.9769 | 47.1365 |
| Average of Profit | 154.117533 | 134.60028 | 118.02049 | 50.0446 | 87.5348 | 129.8065 | 105.2095 | 39.7687 |

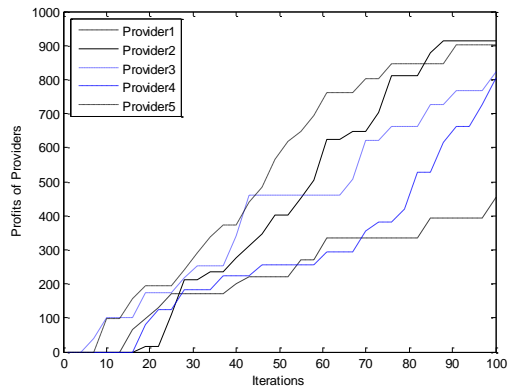

**Fig. 3** Profits of providers using non-competition-based pricing

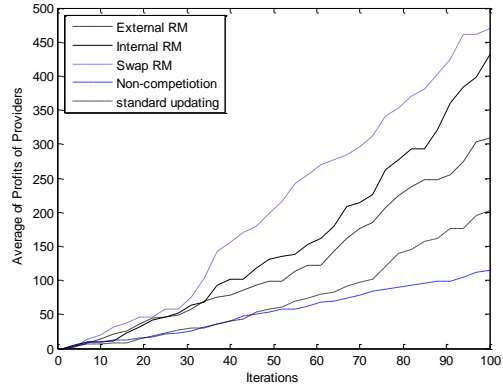

**Fig. 4** Comparison of regret minimization based pricing strategy with non-competition-based pricing and the proposed approach in [42]

Regarding Fig.2, the following results are concluded. Firstly, the profits of providers in swap regret are less than the others, and the profits in external regret are the greatest in almost all iterations. Secondly, in swap regret the upward changes in the profits of providers are less than two other types of regrets. The figures are summarized in Table4. Additionally, Table4 includes the comparison of the proposed pricing approach with two studied pricing policies [14, 26]. The average values of the profit of the providers in some special steps and the average of the values are inserted; the table entries accurately confirm the mentioned points above. In Table4, comparison of the proposed updating technique for probability of strategies of players (Eq.11) with the one in [41] is depicted as well; it can be conveniently observed that our proposed technique has better profits comparing with the standard regret minimization updating techniques (external, internal and swap). Basically, the technique of [41] update the probabilities of strategies in each phase while in Eq.11, the update of probabilities is performed based on the experienced regret. Compared results of our approach with the proposed stochastic game in [14] represent effectiveness of our approach; in [14], SaaS providers in the considered oligopoly have different pricing policies based on the resource types and the benefits that each type has. In our approach, this concept is generalized by $α_i$, as per unit benefit of virtual resources in provider $i$, which makes the approach more applicable. Besides, the price policies in [14] quickly converge to the equilibrium after 10 iterations which cause the same values in rows of Table4. Since in [26], the pricing problem is modeled as a generalized Nash equilibrium problem (GNEP), providers always consider the users and their cost constraints when choosing a pricing strategy. As depicted in Table4 our proposed pricing approach have better profits for the providers in comparison with [26]. In the non-competition-based pricing approach, pricing strategies of providers are independent of the other providers' strategies. In other words, each provider tries to improve its profit without considering other providers' prices. Our competition-based pricing approach of this research is compared with a non-competition-based one (depicted in Table4); from Fig.2 and Fig.3, it can be observed that the profit growth of providers in our approach outperforms the non-competition-based pricing policy.

The profits of providers using Random pricing strategy are added to the Table4 as a benchmark to evaluate the proposed approach. As depicted in Table4, our approach is highly dominating this benchmark.

Additionally, the comparison of profits of 30 providers in a Cloud marketplace is depicted in Fig.4; the average profit of providers per each request are depicted. Regarding the results of this evaluation shows the superiority of the proposed approach to the two other approaches.

It can be observed from Fig.4 that the providers' profits have totally upward trends in much iteration. In [41], standard updating approach is applied to modify the probabilities of the strategies without noting their previous values; the previous behaviors of providers are disregarded in the standard updating approaches introduced in [41]. The results of Fig.4 are also presented in Table4 in details, as well.

In the next experiment, the evolution of regrets of providers while achieving correlated equilibrium is compared in three different regret minimization techniques. Fig.5 depicts the results; as it is expected, the graph shows that the regret is decreasing while the game proceeds. Unlike the other regret techniques, the regrets of external regret technique are initially high, but they slightly decrease. Besides, external regret minimization reaches the equilibrium before both other techniques.

Finally, the probability convergence of the strategy set of providers is studied. It is expected that the evolutions of the probabilities of strategies reach a fixed value while the convergence occurs. Since SaaS providers have a great number of strategies, as discussed in 5.1, the evolutions of a few number of strategies are depicted in Fig.6; here, 10 strategies are supposed. Each line in the figures shows the evolution of the probability of an individual strategy of the provider. It can be observed from Fig.6 that although the convergence of probabilities is obtained in all regret techniques, the external regret reaches faster in comparison with two others. Besides, it can be concluded that although regret minimization causes improvements in providers' decision making, the way probabilities are updated has considerable impacts as well.

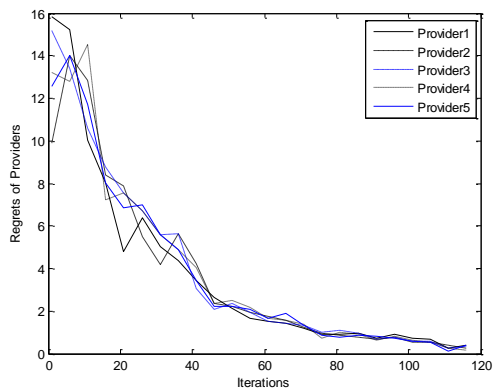

(a)

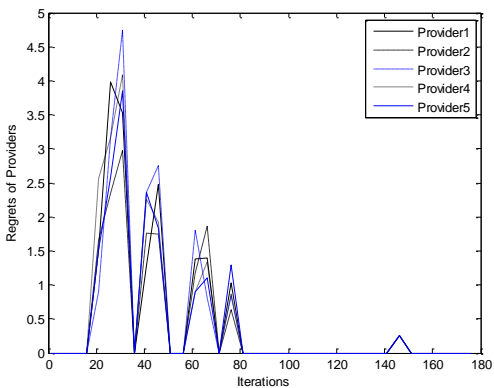

(b)

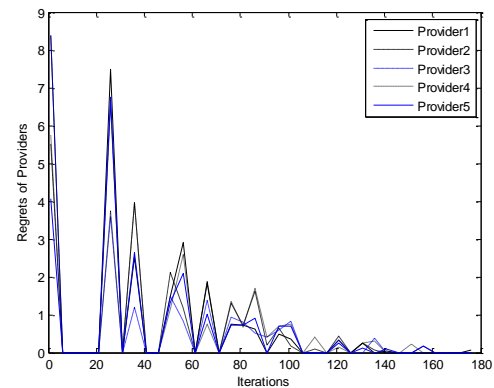

(c)

**Fig. 5** Regrets of providers **(a)** External regret minimization **(b)** Internal regret minimization **(c)** Swap regret minimization

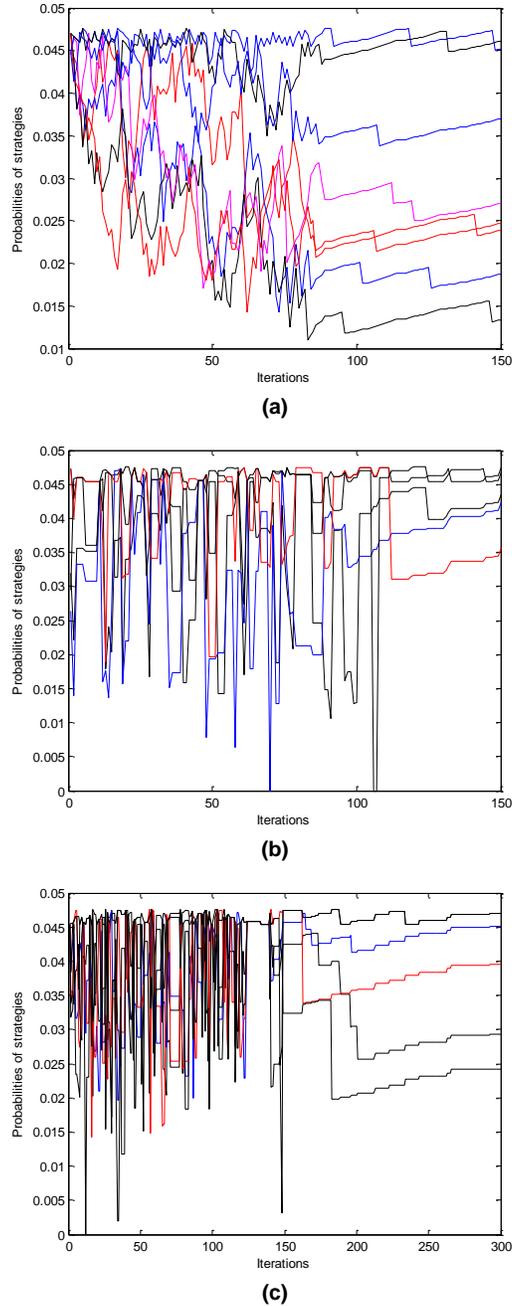

**Fig. 6** The probabilities of 10 number of strategies of a provider in the game **(a)** External regret minimization **(b)** Internal regret minimization **(c)** Swap regret minimization

### 5.3 *Validating the simulation model*

Cost containment is one of the most prominent concerns of SaaS providers; thus, the return on investment of SaaS providers has been discussed in this section. The variations of the investments have been investigated for four providers. The Cloud providers' investments are reported from less than $10M to more than $1B [47]; in this experiment, part of this amount is considered as the investments on the applications, i.e. ($12000~$17000).

Although Fig.7 shows the decline in the performance of the investment in the early rounds, the return on investment is observed in next periods of experiments. Regarding the inequality of Eq.13, it is expected that the profit of provider *i* increases after at most $10 \times L_i \times n$ times. As $n=4$, and $E(\omega_i)=0.1$, providers obtain their investments before $80^{th}$ iteration.

Although the users' preferences have not been considered here, they will surely obtain some benefits from the providers' optimized offers.

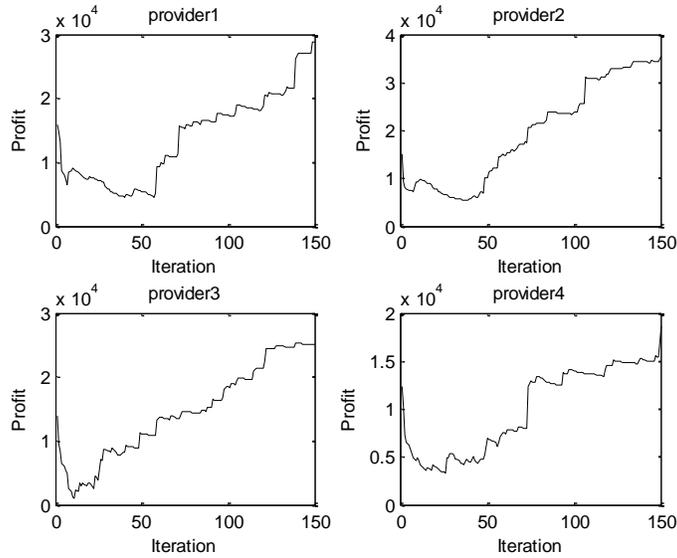

**Fig. 7** Return on investments of SaaS providers in form of profit evolution

The model presented in this article adopts the assumptions and methodology of Cloud models [19, 23, 46, 47], besides, it suffers the same limitations. Adopting more realistic assumptions tends to increase the worthiness of our proposed pricing model. Indeed, if we consider the users' preferences with their limited knowledge of environment, and the IaaS providers' preferences with their limited infrastructure capabilities, our presented pricing mechanism could be a further realistic.

### 5.3.1 *Scalability assessment*

The proposed approach can scale to a realistic size of the Cloud marketplace within an acceptable time limit. On a MacBook Pro with Core 2 Duo processor running at 2.4GHz with 4GB RAM, the experiments are performed with 15 SaaS providers corresponding to the largest number of providers in the current markets and the number of the requests exponentially increases to $10^4$. Although the size of the state space of the market exponentially grows the proposed approach complexity does not increase.

Besides, to validate the reality of the proposed approach on a Cloud marketplace, the experiments are run in a larger scale with a higher number of SaaS providers and requests; the experiments include 15 SaaS providers with request from $10^2$ to $10^4$. The number of applications in each provider is assumed from 100 to 500. The other parameters of the experiment are the same as the previous experiments. Both small and large scale experiments have the same trends; i.e. the least profits belong to swap regret and the best occurs in external regret. The results are summarized in Table5. No figure is reported in this section as depicting 15 vectors belonged to SaaS providers is confusing. Approximate number of required iterations for convergence are computed. As non-competition based pricing approach neglects others players' behaviors without optimizing pricing strategies, it does not converge to any point. The other methods converge to the Equilibrium point after processing less than 10% of requests.

TABLE 5
RESULTS OF EXPERIMENTS IN A LARGER SCALE

| Different R.M. Techniques | External R.M. | Internal R.M. | Swap R.M. |
|---|---|---|---|
| Number of required iterations for convergence | 830 | 790 | 960 |
| Average of Profit | 5080.959 | 4506.303 | 3962.221 |

### 6. CONCLUSION

The Cloud computing environment has recently been considered as a marketplace for providing different services. There are numerous SaaS providers in this market, which might compete or cooperate with each other. In this research the competition of these entities are studied in a way to attract as much as users. Providers compete to set prices of their Cloud applications simultaneously. The competition is modeled as an incomplete-information game, equipped with proposed regret minimization techniques to overcome the information shortage. Players of the game are SaaS providers; the strategies are their pricing policies, and the profits are the estimated utility that providers may obtain by providing the requests. The comparison of efficiency of external, internal and swap regret minimization techniques are studied in the game; each of these techniques follows an individual process to recommend a strategy to players to compute their regret for not choosing the recommended strategy. In each iteration, the players update the distribution of their strategies, which enables them to learn from their past behaviors. In the experiments, the performance of different regret minimization techniques has been evaluated, and the correlated equilibrium of the game is investigated as well. The approach is compared with some different competition-based pricing approaches and the results show a better efficiency of our proposed approach. Finally, since cost saving is one reason that Cloud computing technology is adopted as an alternative to the existing IT solutions, the consequences of cost challenges in the proposed approach are investigated as well. The experiments denote that the proposed approach leads to promising improvements in SaaS providers' return on investment and decision makings on their pricing policies.

There are some future research directions that can be further studied; such as the variety of Cloud services in different time slots and the cooperation of providers may conclude more efficient pricing strategies. There are some limitations in our research; such as lack of open data set related to our work, the current method cannot be verified in the real data.